\documentclass[twocolumn]{jpsj3}

\usepackage{color}
\usepackage{bm}

\title{%
Finite-Size Scaling Analysis of the Conductivity
of Dirac Electrons\\
on a Surface of Disordered Topological Insulators
}

\author{%
Yositake Takane
}

\inst{%
Department of Quantum Matter, Graduate School of Advanced Sciences of Matter,\\
Hiroshima University, Higashihiroshima, Hiroshima 739-8530, Japan
}

\recdate{ \hspace{50mm} }

\abst{%
Two-dimensional (2D) massless Dirac electrons appear on a surface of
three-dimensional topological insulators.
The conductivity of such a 2D Dirac electron system is studied for
strong topological insulators
in the case of the Fermi level being located at the Dirac point.
The average conductivity $\langle\sigma\rangle$ is numerically calculated
for a system of length $L$ and width $W$ under the periodic or antiperiodic
boundary condition in the transverse direction,
and its behavior is analyzed by applying a finite-size scaling approach.
It is shown that $\langle\sigma\rangle$ is minimized at the clean limit,
where it becomes scale-invariant
and depends only on $L/W$ and the boundary condition.
It is also shown that once disorder is introduced,
$\langle\sigma\rangle$ monotonically increases with increasing $L$.
Hence, the system becomes a perfect metal in the limit of $L \to \infty$
except at the clean limit,
which should be identified as an unstable fixed point.
Although the scaling curve of $\langle\sigma\rangle$ strongly depends on $L/W$
and the boundary condition near the unstable fixed point,
it becomes almost independent of them with increasing $\langle\sigma\rangle$,
implying that it asymptotically obeys a universal law.
}

\begin{document}
\maketitle

\section{Introduction}

Three-dimensional topological insulators host
low-energy electron states on their surfaces.~\cite{fu,moore,roy,ando-exp}
These surface electrons are called Dirac electrons
as they obey a massless Dirac equation and hence possess
a linear energy dispersion forming a gapless conic structure
(i.e., a Dirac cone) in the reciprocal space.
Topological insulators are classified into either strong topological insulators
(STIs) or weak topological insulators (WTIs).
The behavior of Dirac electrons is different for these two insulators.
Dirac electrons appear on every surface of an STI,
while they appear only on side surfaces in a typical WTI.
Another important difference is that an STI typically has one Dirac cone,
while a WTI has two Dirac cones.

Let us consider electron transport on a surface of topological insulators
in the presence of disorder,
focusing on the STI case in which surface electrons constitute
a two-dimensional (2D) disordered Dirac electron system with one Dirac cone.
The system under consideration is essentially equivalent to
graphene in the presence of only long-range impurities,
which do not induce intervalley scattering.~\cite{shon}
Thus, to consider electron transport on a surface of STIs,
we can utilize theoretical results for graphene~\cite{shon,titov,rycerz,
nomura1,ostrovsky,bardarson1,ryu,nomura2,tworzydlo,schuessler} in addition to
those of studies focused on topological insulators.~\cite{shindou,
goswami,kobayashi1,kobayashi2,kobayashi3}
An important question is
whether the system exhibits a metal-insulator transition.
The simplest way to answer this question is
to analyze the behavior of the average conductivity
by using a finite-size scaling approach.~\cite{abrahams}
Such studies have been reported in Refs.~\citen{bardarson1}
and \citen{nomura2}.
It is demonstrated that the system shows no metal-insulator transition
and becomes a perfect metal in the thermodynamic limit.
This is in marked contrast to the case of ordinary 2D electron systems
with time-reversal symmetry and
strong spin-orbit interaction.~\cite{hikami,markos,asada}

However, there are subtle issues that need further studies.
In the scaling analysis of disordered graphene
under a two-terminal setup,~\cite{bardarson1}
the periodic boundary condition (PBC) is imposed
in the transverse direction of the 2D Dirac electron system.
In realistic samples of an STI, the antiperiodic boundary condition (ABC)
should be imposed to correctly take account of a Berry phase $\pi$ due to
spin-to-surface locking.~\cite{zhang1,zhang2,egger,bardarson2,imura1}
The PBC applies when an external $\pi$ magnetic flux pierces the system
in the longitudinal direction without touching surface states.
Note that the excitation spectrum of the system changes according to
whether the PBC or the ABC is imposed.
Under the PBC, the system has gapless excitations,
while it acquires a finite-size gap under the ABC.
It has been shown that one perfectly conducting channel is stabilized
in the PBC case, while such a special channel does not appear in the ABC case.
A perfectly conducting channel transmits electrons from one end of the system
to the other without backscattering.~\cite{ando1,ando2,nakanishi,ando3,
takane1,takane2,ando4,sakai,zirnbauer}
We expect that its effect on electron transport depends on
whether the width $W$ of the system is larger or smaller than its length $L$.
The influence of a perfectly conducting channel should be smeared
if $W \gg L$, and hence many conducting channels open at the Fermi level.
However, if $L \gg W$, it should dominate electron transport properties.
Is the boundary condition as well as the aspect ratio $L/W$
completely irrelevant in a finite-size scaling analysis?

Electron transport in the presence of disorder has also
been examined in the WTI case.~\cite{ringel,mong,obuse,takane3}
Dirac electrons on a side surface of WTIs are described
by a 2D chain model~\cite{obuse,morimoto,takane3,arita}
consisting of one-dimensional helical channels, each of which is coupled
with its nearest neighbors.
The number of helical channels plays an important role;
if it is odd, the system has gapless excitations, while it acquires
a finite-size gap if it is even.~\cite{ringel,imura2}
Accordingly, the presence or absence of a perfectly conducting channel is
determined by this parity.~\cite{ringel,takane3,yoshimura}
Hence, the parity of the channel number plays a role similar to
that of the boundary condition in the STI case.
The finite-size scaling analysis that takes account of
the parity effect~\cite{takane3} suggests that this affects
the scaling curve of the average conductivity near the clean limit.

In this paper, we numerically study the behavior of the average conductivity
$\langle\sigma\rangle$ in a disordered 2D Dirac electron system
with a single Dirac cone of length $L$ and width $W$,
focusing on the effects of the transverse boundary condition
and the aspect ratio $R \equiv L/W$.
Our attention is restricted to the case of
the Fermi level being located at the Dirac point,
where $\langle\sigma\rangle$ is expected to be minimized.~\cite{shon,titov}
We numerically calculate $\langle\sigma\rangle$
for the system with a fixed $R$ under the PBC or the ABC,
and analyze its behavior using a finite-size scaling approach.
It is shown that $\langle\sigma\rangle$ is minimized at the clean limit,
where it becomes scale-invariant
and depends only on $R$ and the boundary condition.
It is also shown that $\langle\sigma\rangle$ monotonically increases
with increasing $L$ in the presence of disorder.
From these results, we conclude that the system becomes a perfect metal
in the limit of $L \to \infty$ except at the clean limit,
which should be identified as an unstable fixed point.
The scaling curve of $\langle\sigma\rangle$ depends on $R$
and the boundary condition (i.e., the presence/absence of a finite-size gap)
near the unstable fixed point.
However, it becomes almost independent of them
with increasing $\langle\sigma\rangle$,
implying that it asymptotically obeys a universal law.
We set $\hbar = 1$ throughout this paper.

\section{Model and Formulation}

\begin{figure}[bpt]
\begin{center}
\includegraphics[height=3.5cm]{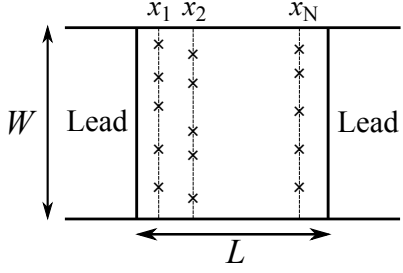}
\end{center}
\caption{Structure of a 2D sample of length $L$ and width $W$
considered in the text, where crosses represent impurities.
The role of the dashed lines is explained at the end of Sect.~2.
}
\end{figure}
Let us consider a disordered 2D Dirac electron system of width $W$
in the $y$-direction being infinitely long in the $x$-direction (see Fig.~1).
We impose either the PBC or the ABC in the $y$-direction.
The region of $L \ge x \ge 0$ is regarded as the sample with disorder,
and the region of $x < 0$ ($x > L$) plays the role of the left (right) lead.
We adopt the following Dirac Hamiltonian:
\begin{align}
       \label{H_eff}
   H
   =  v \left[ \begin{array}{cc}
                 -i\partial_{x}+V(x,y) & -\partial_{y} \\
                 \partial_{y} & i\partial_{x}+V(x,y)
               \end{array} \right] 
\end{align}
with $V(x,y) \equiv V_{\rm imp}(x,y)+U(x)$.
Here, $V_{\rm imp}(x,y)$ represents the disorder potential
arising from random impurities,
while $U(x)$ is included to simulate the setup~\cite{katsnelson}
in which the Fermi level is fixed at the Dirac point ($\epsilon = 0$)
in the sample region, while the left and right leads are deeply doped.
Accordingly, we set $U(x) = 0$ for $L > x > 0$,
and otherwise $U(x) = - U_0$ with $U_0$ being positive and large.
Let us briefly describe the wave functions at an energy $\epsilon$
in the case of $V(x,y) \equiv 0$.
The transverse function is given by
\begin{align}
    \label{eq:transv-f}
  \chi_{m}(y) = \frac{1}{\sqrt{W}}e^{iq_{m}y}
                \left[ \begin{array}{c}
                          a_{m}^{\pm} \\
                          b_{m}^{\pm}
                       \end{array} \right] ,
\end{align}
where $q_{m} = 2\pi m/W$ and
\begin{align}
   \left[ \begin{array}{c}
            a_{m}^{\pm} \\
            b_{m}^{\pm}
          \end{array} \right]
   = \frac{1}{\sqrt{|\pm k_{m}-\frac{\epsilon}{v}|^2+q_{m}^2}}
     \left[ \begin{array}{c}
            iq_{m} \\
            \pm k_{m}-\frac{\epsilon}{v}
          \end{array} \right] .
\end{align}
Here, $m$ is given by
\begin{align}
  m = 0, \pm 1, \pm2, \dots
\end{align}
in the PBC case and
\begin{align}
  m = \pm \frac{1}{2}, \pm \frac{3}{2}, \pm \frac{5}{2}, \cdots
\end{align}
in the ABC case.
For the $m$th mode, the dispersion relation as a function of
the longitudinal wave number $k$
is $\epsilon_{m}(k) = \pm v\sqrt{k^{2}+q_{m}^{2}}$.
In the PBC case, $q_{m}$ vanishes for $m = 0$,
indicating that the system has gapless excitations,
while $m = 0$ is not allowed in the ABC case
so a finite-size gap opens across the Dirac point.
If $|\epsilon| > v|q_{m}|$,
the $m$th mode provides two counterpropagating channels.
The corresponding wave functions are
\begin{align}
      \label{eq:wf-propa}
   \varphi_{m}^{\pm}(x,y)
   = \frac{1}{\sqrt{v_{m}}} \chi_{m}(y)
     e^{\pm ik_{m}x} ,
\end{align}
where $\pm$ specifies the propagating direction,
$k_{m} = \sqrt{(\epsilon/v)^2-q_{m}^2}$,
and the group velocity $v_{m}$ is obtained as $v_{m} = v(vk_{m}/|\epsilon|)$.
If $|\epsilon| < v|q_{m}|$, the $m$th mode provides two evanescent channels.
The corresponding wave functions are obtained
from Eq.~(\ref{eq:wf-propa}) by the replacement of
$k_{m}$ with $i\kappa_{m}$,
where $\kappa_{m} = \sqrt{q_{m}^2-(\epsilon/v)^2}$.
The group velocity has no physical meaning for evanescent channels.
The number of modes should be restricted to a finite number $M$
in actual numerical calculations.
We set $M$ be an odd (even) integer in the PBC (ABC) case,
and then restrict $m$ as $m_{\rm max} \ge m \ge -m_{\rm max}$ with
\begin{align}
  m_{\rm max} = \frac{M-1}{2} .
\end{align}
Here, we define the short length scale $a$ by $a \equiv W/M$,
which is on the order of the spatial resolution of plane waves
limited by the restriction of transverse modes.
This is equivalent to saying that $W$ is an integer multiple of $a$,
\begin{align}
  M \equiv W/a .
\end{align}
We also assume that $L$ is an integer multiple of $a$ and set
\begin{align}
  N \equiv L/a .
\end{align}
The disorder potential $V_{\rm imp}(x,y)$ is assumed to consist
of $\delta$-function-type impurities in the sample region,
\begin{align}
  V_{\rm imp}(x,y)
  = \sum_{p=1}^{N_{\rm imp}}V_{p}a^{2}
                            \delta(x-x_{p})\delta(y-y_{p}) ,
\end{align}
where $V_p$ is the strength of the $p$th impurity located at
$(x,y) = (x_p,y_{p})$ and $N_{\rm imp}$ is the total number of impurities.
We define $\gamma$ in terms of the correlation function
of $V_{\rm imp}(x,y)$ as~\cite{rycerz}
\begin{align}
  \gamma = \frac{1}{v^2}\int_{0}^{L} dx' \int_{0}^{W} dy'
           \langle V_{\rm imp}(x,y)V_{\rm imp}(x',y') \rangle ,
\end{align}
where $\langle \cdots \rangle$ represents the disorder average.
If $V_p$ is assumed to be uniformly distributed within the interval of
$[-V_{0},+V_{0}]$, we find that $\gamma = (1/3)\Gamma$ with
\begin{align}
  \Gamma  = \left(\frac{a}{v}\right)^{2}
            \frac{a^{2}}{LW} N_{\rm imp}V_{0}^{2} .
\end{align}
We use $\Gamma$ to characterize the strength of disorder.

The dimensionless conductance $g$ of a sample with impurities is determined by
the Landauer formula,
\begin{align}
  g = {\rm tr}\{\mib{t}^{\dagger}\mib{t}\} ,
\end{align}
where $\mib{t}$ is the transmission matrix.
As the area of the sample is $W \times L$, the dimensionless conductivity
$\sigma$ is given by
\begin{align}
     \label{eq:sigma}
  \sigma = \frac{L}{W} g .
\end{align}
The transmission matrix for a given impurity configuration is
numerically determined by using the method presented in Ref.~\citen{tamura}.
In this method, the $S$ matrix for the whole system is decomposed into
single-impurity parts and free-propagating parts.
Once they are evaluated, we can construct the $S$ matrix
using a composition law.
If the above method is straightforwardly applied to our system,
the number of single-impurity parts is $N_{\rm imp}$ and
that of free-propagating parts is $N_{\rm imp}+1$.
In actual numerical calculations, we set $N_{\rm imp} = N \times M$.
To save computational time, we employ the following procedure
(see Fig.~1):~\cite{takane3}
randomly choose $N$ points on the $x$-axis in the sample region
such that $L> x_{N} > \dots > x_2 > x_1 >0$ and then randomly place
$M$ impurities on each line of width $W$ at $x = x_i$ ($i = 1,2,\dots,N$).
With this procedure, the number of single-impurity parts is reduced to $N$
while the total number of impurities is $N_{\rm imp}=N \times M$.

\section{Numerical Results}

\begin{figure}[tbp]
\begin{center}
\includegraphics[height=6.5cm]{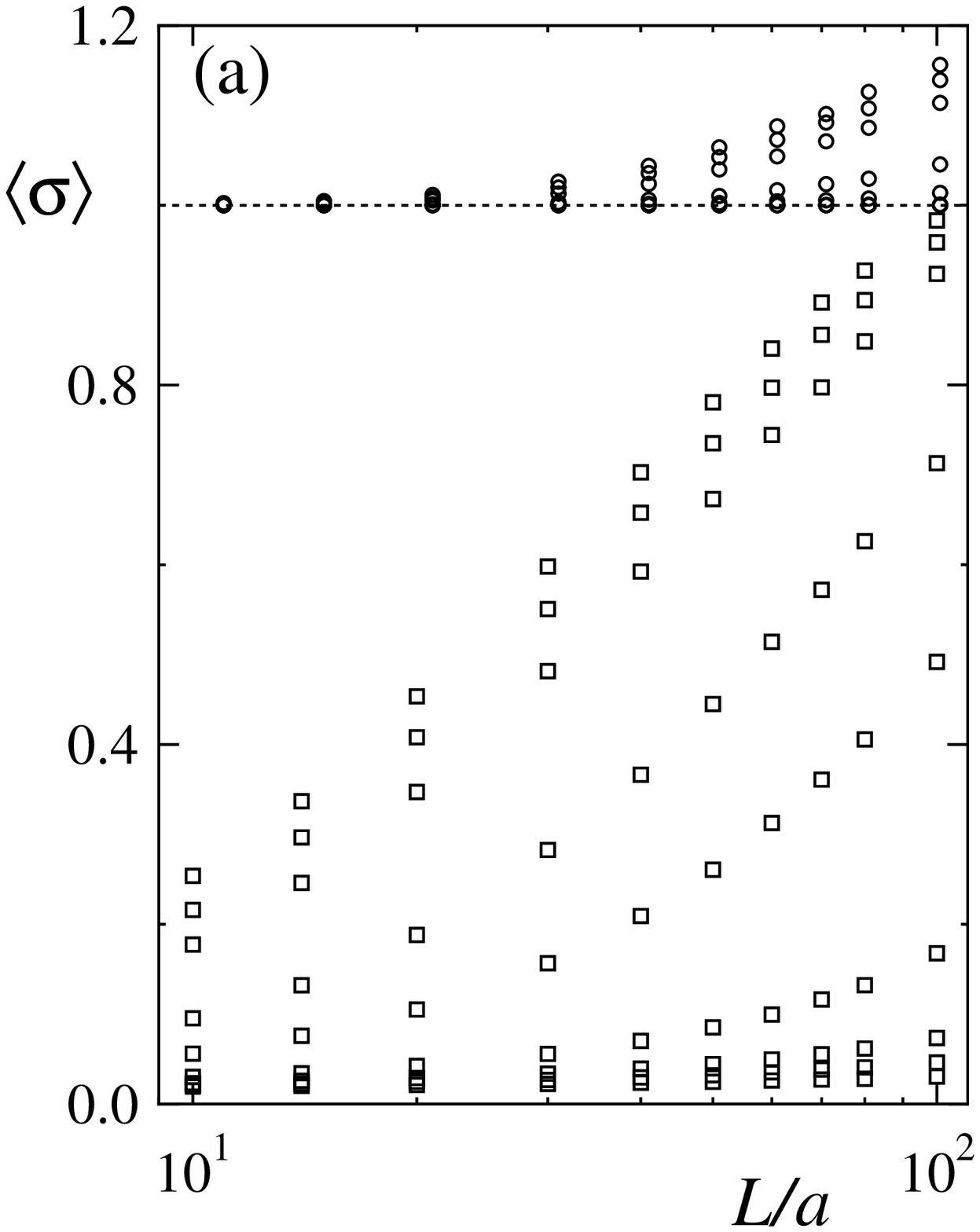}
\end{center}
\begin{center}
\includegraphics[height=6.5cm]{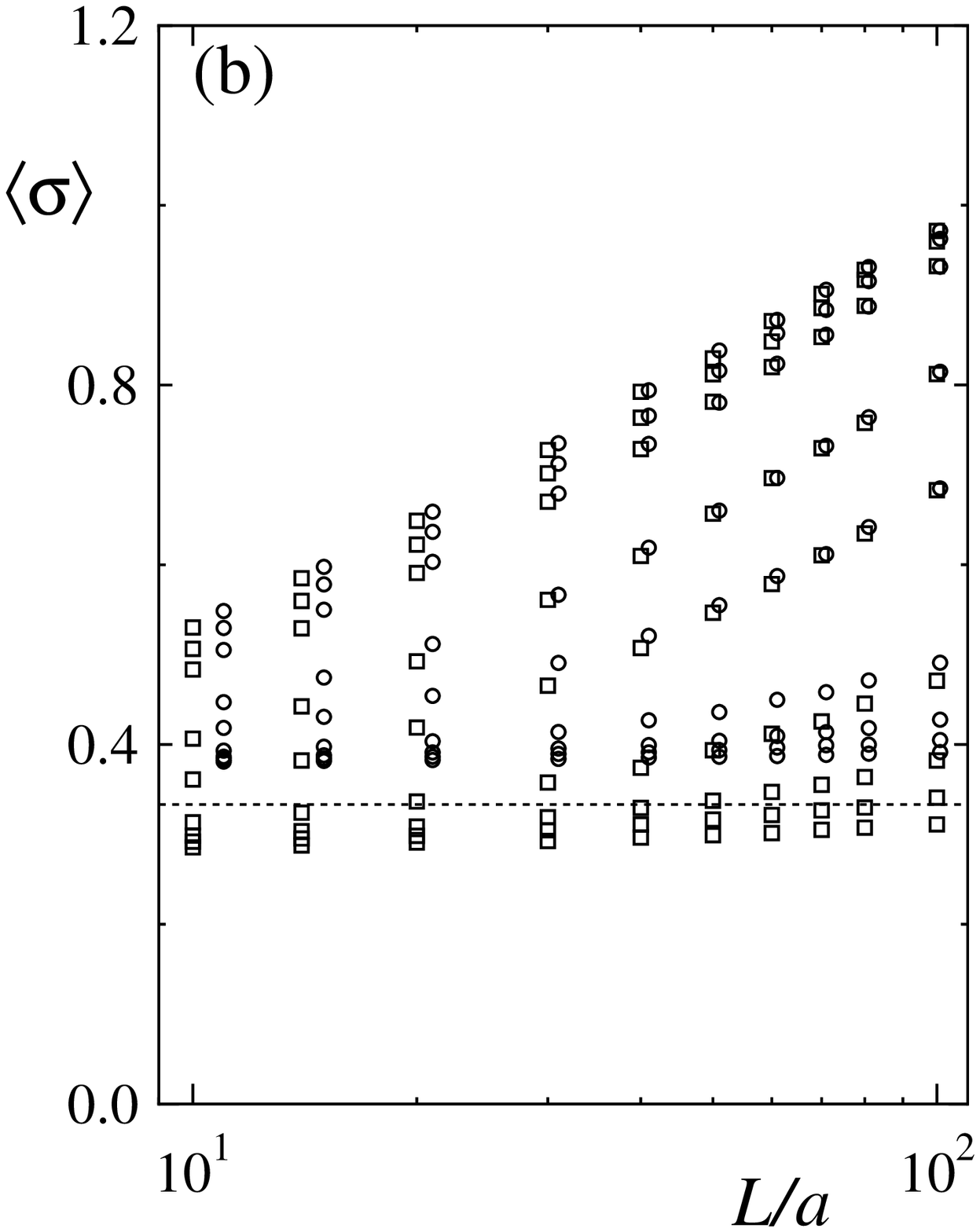}
\end{center}
\caption{Average conductivity in the cases of (a) $R = 1$
and (b) $R = 1/3$ as functions of $L/a$
for $\Gamma = 0.6$, $0.7$, $0.8$, $1.0$, $1.5$, $2.0$, $3.0$, $3.5$,
and $4.0$ from bottom to top.
Circles and squares respectively correspond to the PBC case and the ABC case.
Dotted lines designate $\langle \sigma \rangle = 1$ in (a)
and $\langle \sigma \rangle = 1/3$ in (b).
}
\end{figure}
Now, we present the numerical results of the average conductivity
$\langle\sigma\rangle$ for various system sizes
with a fixed aspect ratio $R \equiv N/M$.
We particularly consider the two cases of $R = 1$ and $R = 1/3$.
In the case of $R = 1$, the system size is varied from
$N \times M = 11 \times 11$ to $201 \times 201$ in the PBC case and
from $10 \times 10$ to $200 \times 200$ in the ABC case.
In the case of $R = 1/3$, the system size is varied from
$N \times M = 11 \times 33$ to $141 \times 423$ in the PBC case
and from $10 \times 30$ to $140 \times 420$ in the ABC case.
As the number of impurities is fixed at $N_{\rm imp} = N \times M$,
the strength of disorder is simply expressed as $\Gamma  = (a/v)^{2}V_{0}^{2}$.
This value is tuned within $\Gamma = 0.6$--$4.0$ by adjusting $V_{0}$.
Actually, we examine the cases with $\Gamma = 0.6$, $0.7$, $0.8$, $1.0$,
$1.5$, $2.0$, $3.0$, $3.5$, and $4.0$.
When performing the ensemble average,
the number of samples, $N_{\rm sam}$, is set larger than $5000$
so that the relative uncertainty $\Delta\sigma/\langle\sigma\rangle$
is smaller than $0.005$ at each data point,
where $\Delta\sigma \equiv ({\rm var}\{\sigma\}/N_{\rm sam})^{1/2}$.

The $L$ dependence of $\langle \sigma \rangle$ is shown in Fig.~2
within the interval of $101 \ge L/a \ge 10$.
Clearly, $\langle \sigma \rangle$ increases with increasing $\Gamma$,
and it also increases with increasing $L$
although the rate of increase becomes very small at small $\Gamma$.
We observe that $\langle \sigma \rangle$ behaves differently
depending on $R$ and/or the boundary condition.
Hence, the following four cases are separately considered below:
the cases of $R = 1$ under the PBC and the ABC,
and the cases of $R = 1/3$ under the PBC and the ABC.
Notably, $\langle \sigma \rangle$ is always larger than unity
in the case of $R = 1$ under the PBC [see Fig.~2(a)],
which results from the fact that
one perfectly conducting channel is stabilized under the PBC.
The presence of a perfectly conducting channel ensures that
$\langle g \rangle > 1$,
and this directly accounts for the behavior
since $\langle \sigma \rangle = \langle g \rangle$ in this case.
In the case of $R = 1/3$ under the PBC, the average conductivity is
also bounded below as $\langle \sigma \rangle > 1/3$
since $\langle \sigma \rangle = (1/3)\langle g \rangle$.
Such a simple lower bound is absent under the ABC.
Thus, we can conclude that the boundary-condition dependence of
$\langle \sigma \rangle$ is mainly induced by
the presence or absence of a perfectly conducting channel.

\begin{figure}[bpt]
\begin{center}
\includegraphics[height=7.8cm]{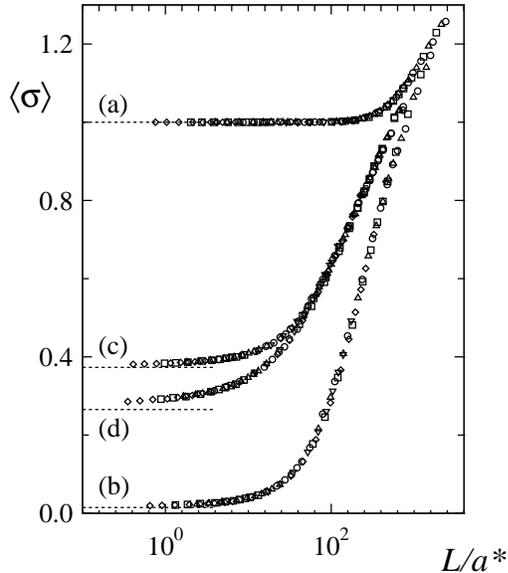}
\end{center}
\caption{One-parameter scaling plot of the average conductivity
for (a) the case with $R = 1$ under the PBC, (b) that under the ABC,
(c) the case with $R = 1/3$ under the PBC, and (d) that under the ABC.
Here, the data sets for $\Gamma = 0.6$, $0.7$, $0.8$, $1.0$, $1.5$, $2.0$,
$3.0$, $3.5$, and $4.0$ are used.
Dotted lines indicate
the clean-limit values of the conductivity (see the text).
}
\end{figure}
As demonstrated in Fig.~3, in each of the four cases, the data sets for
$\Gamma = 0.6$--$4.0$ collapse onto one scaling curve
upon shifting the data horizontally,
where the new length scale $a^{*}$ is separately determined for each data set.
Note that $a^{*}$ decreases with increasing $\Gamma$,
playing a role similar to the mean free path.
Figure~3 clearly shows that the behaviors of $\langle \sigma \rangle$
in the four cases are different for small $L/a^{*}$.
However, we observe that the corresponding four curves merge into
one scaling curve in the region of $L/a^{*} \gtrsim 500$.
This implies that the scaling curve of $\langle \sigma \rangle$
asymptotically obeys a universal law
in the long-length (or strong-disorder) regime of large $L/a^{*}$,
although it exhibits nonuniversal behavior
(i.e., significant dependence on $R$ and the boundary condition)
in the short-length (or weak-disorder) regime of small $L/a^{*}$.

We also observe that the average conductivity decreases toward
a limiting value with decreasing $L/a^*$.
Since $\langle \sigma \rangle$ decreases with decreasing $\Gamma$,
it is reasonable to identify this limiting value
with the conductivity in the clean limit,
which is denoted by $\sigma_{\rm cl}$ hereafter.
Note that $\sigma_{\rm cl}$ does not depend on the system size
and is determined by only $R$ and the boundary condition.
This follows from the fact that the conductance $g_{\rm cl}$ at the clean limit
is analytically given by~\cite{katsnelson}
\begin{align}
  g_{\rm cl}
  = \sum_{m_{\rm max} \ge |m|}\cosh^{-2}\left( q_{m}L \right)
\end{align}
with $q_{m}L$ being rewritten as $q_{m}L = 2\pi m R$.~\cite{comment1}
Note that the term with $m=0$, which is allowed only in the PBC case,
corresponds to a perfectly conducting channel.
We can numerically evaluate $\sigma_{\rm cl}$ from this expression.
The dotted lines in Fig.~3 indicate the resulting values in the four cases.
Figure~3 supports the reasoning that $\langle \sigma \rangle$
approaches $\sigma_{\rm cl}$ with decreasing $L/a^{*}$ in each case.

Let us consider the behavior of the scaling function $\beta$ defined by
\begin{align}
  \beta\left(\langle \sigma \rangle\right)
  = \frac{d{\rm ln}\langle \sigma \rangle}{d{\rm ln}L} .
\end{align}
This is determined from the data shown in Fig.~3,
and the result for the case with $R = 1$ is shown in Fig.~4.
As $\langle \sigma \rangle$ monotonically increases with increasing $L/a^{*}$,
$\beta$ is always positive.
If $\langle \sigma \rangle$ becomes sufficiently large,
we expect that $\beta$ obeys~\cite{nomura2}
\begin{align}
      \label{eq:beta-universal}
  \beta\left(\langle \sigma \rangle\right)
  = \frac{1}{\pi\langle \sigma \rangle} ,
\end{align}
in accordance with the weak antilocalization theory.~\cite{suzuura}
Unfortunately, the regime of such a large $\langle \sigma \rangle$
is not examined in this study owing to the limitation of the computing time.
In the opposite regime of $\langle \sigma \rangle$ close to $\sigma_{\rm cl}$,
$\beta$ becomes small because the rate of increase in $\langle \sigma \rangle$
is strongly suppressed near the clean limit.
Precisely at the clean limit where $\langle \sigma \rangle = \sigma_{\rm cl}$,
$\beta$ vanishes since $\sigma_{\rm cl}$ is independent of $L$.
We conclude that $\beta(\langle \sigma \rangle)$ is always positive
except at the clean limit,
which should be identified as an unstable fixed point.
The qualitative features of $\beta$ argued above apply
irrespective of $R$ and the boundary condition
although the $\langle \sigma \rangle$ dependence of $\beta$ is
strongly modified at small $\langle \sigma \rangle$
if $R$ and the boundary condition are changed.
However, $\beta$ becomes almost independent of them
for sufficiently large $\langle \sigma \rangle$.
This implies that the universal behavior of Eq.~(\ref{eq:beta-universal})
asymptotically manifests itself with increasing $\langle \sigma \rangle$.
\begin{figure}[bpt]
\begin{center}
\includegraphics[height=4.8cm]{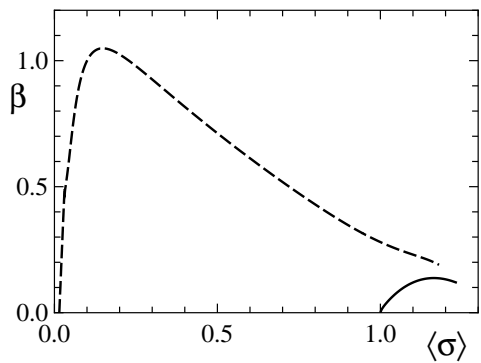}
\end{center}
\caption{Scaling function for the case with $R = 1$
determined from the data shown in Fig.~3,
where the solid line (dashed line) represents
the result under the PBC (ABC).
}
\end{figure}

\section{Summary and Discussion}

We have studied the conductivity of a 2D Dirac electron system
with one Dirac cone in the two-terminal setup,
in which the Fermi level is fixed at the Dirac point in the sample region,
while the left and right leads are deeply doped.
The average conductivity $\langle \sigma \rangle$ is numerically calculated
for the system of length $L$ and width $W$ with a fixed aspect ratio
$R \equiv L/W$ under the periodic or antiperiodic
boundary condition in the transverse direction,
where the periodic (antiperiodic) boundary condition directly results in
the absence (presence) of a finite-size gap.
It is shown that $\langle \sigma \rangle$ monotonically increases
with increasing $L$ except at the clean limit,
where the conductivity becomes scale-invariant
but strongly depends on $R$ and the boundary condition.
The scaling function $\beta(\langle \sigma \rangle)$
is determined from the resulting $\langle \sigma \rangle$.
The behavior of $\beta$ is similar to that found in Ref.~\citen{bardarson1}.
However, we uncover two features not demonstrated in Ref.~\citen{bardarson1}:
the clean limit corresponds to the unstable fixed point at which
$\beta(\langle \sigma \rangle)$ vanishes,~\cite{schuessler}
and $\beta(\langle \sigma \rangle)$ strongly depends on $L/W$ and
the boundary condition near the unstable fixed point, although it
asymptotically obeys a universal law with increasing $\langle \sigma \rangle$.

The scaling function $\beta(\langle \sigma \rangle)$ has also been determined
from the data of $\langle \sigma \rangle$
obtained by a linear response calculation in the momentum space
while varying the Fermi level.~\cite{nomura2}
The resulting $\beta(\langle \sigma \rangle)$ behaves differently from
that demonstrated in Fig.~4;
it monotonically decreases with increasing $\langle \sigma \rangle$
in accordance with the universal behavior, i.e., Eq.~(\ref{eq:beta-universal}).
As this calculation treats the system with no electrode
away from the Dirac point, the entire setup is completely different from
that assumed in the present study.
Therefore, we need not seriously consider the difference.
However, note that $\beta(\langle \sigma \rangle)$ is not determined
in the regime of small $\langle \sigma \rangle$ in Ref.~\citen{nomura2}.
If such a regime were also analyzed within the same approach,
its own setup might play a role, resulting in nonuniversal behavior
that is different from that demonstrated in Fig.~4.
Roughly speaking, Ref.~\citen{nomura2} mainly focuses on the universal behavior
of $\beta(\langle \sigma \rangle)$, in contrast with
that in ordinary 2D electron systems with time-reversal symmetry and
strong spin-orbit interaction,
while the present study focuses on how $\langle \sigma \rangle$ approaches
the universal regime starting from the clean limit
under a particular setup.

\section*{Acknowledgment}

This work was supported by JSPS KAKENHI Grant Number 15K05130.


\begin{thebibliography}{99}

\bibitem{fu} L. Fu, C. L. Kane, and E. J. Mele,
Phys. Rev. Lett. {\bf 98}, 106803 (2007).

\bibitem{moore} J. E. Moore and L. Balents,
Phys. Rev. B {\bf 75}, 121306 (2007).

\bibitem{roy} R. Roy, Phys. Rev. B {\bf 79}, 195322 (2009).

\bibitem{ando-exp} Y. Ando, J. Phys. Soc. Jpn. {\bf 82}, 102001 (2013).

\bibitem{shon} N. H. Shon and T. Ando,
J. Phys. Soc. Jpn. {\bf 67}, 2421 (1998).

\bibitem{titov} M. Titov, EPL {\bf 79}, 17004 (2007).

\bibitem{rycerz} A. Rycerz, J. Tworzyd{\l}o, and C. W. J. Beenakker,
EPL {\bf 79}, 57003 (2007).

\bibitem{nomura1} K. Nomura and A. H. MacDonald,
Phys. Rev. Lett. {\bf 98}, 076602 (2007).

\bibitem{ostrovsky} P. M. Ostrovsky, I. V. Gornyi, and A. D. Mirlin,
Phys. Rev. Lett. {\bf 98}, 256801 (2007).

\bibitem{bardarson1} J. H. Bardarson, J. Tworzyd{\l}o, P. W. Brouwer,
and C. W. J. Beenakker, Phys. Rev. Lett. {\bf 99}, 106801 (2007).

\bibitem{ryu} S. Ryu, C. Mudry, H. Obuse, and A. Furusaki,
Phys. Rev. Lett. {\bf 99}, 116601 (2007).

\bibitem{nomura2} K. Nomura, M. Koshino, and S. Ryu,
Phys. Rev. Lett. {\bf 99}, 146806 (2007).

\bibitem{tworzydlo} J. Tworzyd{\l}o, C. W. Groth, and C. W. J. Beenakker,
Phys. Rev. B {\bf 78}, 235438 (2008).

\bibitem{schuessler} A. Schuessler, P. M. Ostrovsky, I. V. Gornyi,
and A. D. Mirlin, Phys. Rev. B {\bf 79}, 075405 (2009).

\bibitem{shindou} R. Shindou and S. Murakami,
Phys. Rev. B {\bf 79}, 045321 (2009).

\bibitem{goswami} P. Goswami and S. Chakravarty,
Phys. Rev. Lett. {\bf 107}, 196803 (2011).

\bibitem{kobayashi1} K. Kobayashi, T. Ohtsuki, and K.-I. Imura,
Phys. Rev. Lett. {\bf 110}, 236803 (2013).

\bibitem{kobayashi2} K. Kobayashi, T. Ohtsuki, K.-I. Imura, and I. F. Herbut,
Phys. Rev. Lett. {\bf 112}, 016402 (2014).

\bibitem{kobayashi3} K. Kobayashi, Y. Yoshimura, K.-I. Imura, and T. Ohtsuki,
Phys. Rev. B {\bf 92}, 235407 (2015).

\bibitem{abrahams} E. Abrahams, P. W. Anderson, D. C. Licciardello,
and T. V. Ramakrishnan, Phys. Rev. Lett. {\bf 42}, 673 (1979).

\bibitem{hikami} S. Hikami, A. I. Larkin, and Y. Nagaoka,
Prog. Theor. Phys. {\bf 63}, 707 (1980).

\bibitem{asada} Y. Asada, K. Slevin, and T. Ohtsuki,
Phys. Rev. B {\bf 70}, 035115 (2004).

\bibitem{markos} P. Marko\v{s} and L. Schweitzer,
J. Phys. A {\bf 39}, 3221 (2006).

\bibitem{zhang1} Y. Zhang, Y. Ran, and A. Vishwanath,
Phys. Rev. {\bf 79}, 245331 (2009).

\bibitem{zhang2} Y. Zhang and A. Vishwanath,
Phys. Rev. Lett. {\bf 105}, 206601 (2010).

\bibitem{egger} R. Egger, A. Zazunov, and A. Levy Yeyati,
Phys. Rev. Lett. {\bf 105}, 136403 (2010).

\bibitem{bardarson2} J. H. Bardarson, P. W. Brouwer, and J. E. Moore,
Phys. Rev. Lett. {\bf 105}, 156803 (2010).

\bibitem{imura1} K.-I. Imura, Y. Takane, and A. Tanaka,
Phys. Rev. B {\bf 84}, 195406 (2011).

\bibitem{ando1} T. Ando and T. Nakanishi,
J. Phys. Soc. Jpn. {\bf 67}, 1704 (1998).

\bibitem{ando2} T. Ando, T. Nakanishi, and R. Saito,
J. Phys. Soc. Jpn. {\bf 67}, 2857 (1998).

\bibitem{nakanishi} T. Nakanishi and T. Ando,
J. Phys. Soc. Jpn. {\bf 68}, 561 (1999).

\bibitem{ando3} T. Ando and H. Suzuura,
J. Phys. Soc. Jpn. {\bf 71}, 2753 (2002).

\bibitem{takane1} Y. Takane,
J. Phys. Soc. Jpn. {\bf 73}, 1430 (2004).

\bibitem{takane2} Y. Takane,
J. Phys. Soc. Jpn. {\bf 73}, 2366 (2004).

\bibitem{ando4} T. Ando,
J. Phys. Soc. Jpn. {\bf 75}, 054701 (2006).

\bibitem{sakai} H. Sakai and Y. Takane,
J. Phys. Soc. Jpn. {\bf 75}, 054711 (2006).

\bibitem{zirnbauer} M. R. Zirnbauer,
Phys. Rev. Lett. {\bf 69}, 1584 (1992).

\bibitem{ringel} Z. Ringel, Y. E. Kraus, and A. Stern,
Phys. Rev. B {\bf 86}, 045102 (2012).

\bibitem{mong} R. S. K. Mong, J. H. Bardarson, and J. E. Moore,
Phys. Rev. Lett. {\bf 108}, 076804 (2012).

\bibitem{obuse} H. Obuse, S. Ryu, A. Furusaki, and C. Mudry,
Phys. Rev. B {\bf 89}, 155315 (2014).

\bibitem{takane3} Y. Takane,
J. Phys. Soc. Jpn. {\bf 83}, 103706 (2014).

\bibitem{morimoto} T. Morimoto and A. Furusaki,
Phys. Rev. B {\bf 89}, 035117 (2014).

\bibitem{arita} T. Arita and Y. Takane,
J. Phys. Soc. Jpn. {\bf 83}, 124716 (2014).

\bibitem{imura2} K.-I. Imura, M. Okamoto, Y. Yoshimura, Y. Takane,
and T. Ohtsuki, Phys. Rev. B {\bf 86}, 245436 (2012).

\bibitem{yoshimura} Y. Yoshimura, A. Matsumoto, Y. Takane, and K.-I. Imura,
Phys. Rev. B {\bf 88}, 045408 (2013).

\bibitem{katsnelson} M. I. Katsnelson, Eur. Phys. J. B {\bf 51}, 157 (2006).

\bibitem{tamura} H. Tamura and T. Ando, Phys. Rev. B {\bf 44}, 1792 (1991).

\bibitem{comment1} Precisely speaking, the $M$ dependence of the cutoff
$m_{\rm max}$ yields a size dependence of $g_{\rm cl}$.
However, it is extremely small and can be safely neglected.

\bibitem{suzuura} H. Suzuura and T. Ando,
Phys. Rev. Lett. {\bf 89}, 266603 (2002).




\end{thebibliography}
\end{document}